\documentclass[aps,pra,twocolumn,nofootinbib,superscriptaddress,letterpaper]{revtex4-1}

\usepackage[utf8]{inputenc}
\usepackage[english]{babel}

\usepackage{graphicx}
\usepackage{siunitx}
\usepackage{eurosym}
\usepackage{xcolor}
\usepackage{verbatim}

\usepackage{hyperref}
\hypersetup{
    colorlinks,
    citecolor=blue,
    filecolor=blue,
    linkcolor=black,
    urlcolor=blue
}

\usepackage{amsmath}
\usepackage{amsfonts}
\usepackage{amssymb}

\usepackage{natbib}
\usepackage[version=3]{mhchem}

\begin{document}

%%%%%%%%%%%%%%%%%%%%%%%%%%%%%%%%%%%%%%%%%%%%%
\author{Carlos Saavedra}
 \altaffiliation{Current address: Department of Chemistry, University of Wisconsin, Madison, USA}
 \affiliation{Institute of Applied Physics, University of Bonn, Germany}
 \affiliation{División de Ciencias e Ingenierias, Universidad de Guanajuato, México}

\author{Deepak Pandey}
 \altaffiliation{Current address: Menlo Systems GmbH, Am Klopferspitz 19a, 82152, Martinsried, Germany}
 \affiliation{Institute of Applied Physics, University of Bonn, Germany}

\author{Wolfgang Alt}
 \affiliation{Institute of Applied Physics, University of Bonn, Germany}

\author{Dieter Meschede}
 \affiliation{Institute of Applied Physics, University of Bonn, Germany}

\author{Hannes Pfeifer}
 \affiliation{Institute of Applied Physics, University of Bonn, Germany}

%%%%%%%%%%%%%%%%%%%%%%%%%%%%%%%%%%%%%%%%%%%%%
\title{A fiber Fabry-Perot cavity based spectroscopic gas sensor}

%%%%%%%%%%%%%%%%%%%%%%%%%%%%%%%%%%%%%%%%%%%%%
\date{\today}

%%%%%%%%%%%%%%%%%%%%%%%%%%%%%%%%%%%%%%%%%%%%%
\begin{abstract}
Optical spectroscopic sensors are powerful tools for analysing gas mixtures in industrial and scientific applications. Whilst highly sensitive spectrometers tend to have a large footprint, miniaturized optical devices usually lack sensitivity or wideband spectroscopic coverage. By employing a widely tunable, passively stable fiber Fabry-Perot cavity (FFPC), we demonstrate an absorption spectroscopic device that continuously samples over several tens of terahertz. Both broadband scans using cavity mode width spectroscopy to identify the spectral fingerprints of analytes and a fast, low-noise scan method for single absorption features to determine concentrations are exemplary demonstrated for the oxygen A-band. The novel scan method uses an injected modulation signal in a Pound-Drever-Hall feedback loop together with a lock-in measurement to reject noise at other frequencies. The FFPC-based approach provides a directly fiber coupled, extremely miniaturized, light-weight and robust platform for analyzing small analyte volumes that can straightforwardly be extended to sensing at different wavelength ranges, liquid analytes and other spectroscopic techniques with only little adjustments of the device platform. 
\end{abstract}

%%%%%%%%%%%%%%%%%%%%%%%%%%%%%%%%%%%%%%%%%%%%%
\maketitle

%%%%%%%%%%%%%%%%%%%%%%%%%%%%%%%%%%%%%%%%%%%%%
\section{Introduction}
Optical absorption spectroscopy is a key tool for analysing the chemical composition of matter since the early days of optics \cite{wollaston1802xii,fraunhofer1817bestimmung,kirchhoff1860fraunhofer}. Its sensitivity depends on the optical depth of the analyte, which can be strongly enhanced by enclosing it in an optical cavity. Cavity enhanced absorption spectroscopy using cavity ring-downs \cite{zalicki1995cavity,berden2000cavity}, mode width \cite{cygan2013cavity}, or more elaborate techniques like noise-immune cavity-enhanced optical heterodyne molecular spectroscopy (NICE-OHMS) \cite{foltynowicz2008noise} have found applications in many fields of research, in particular in gas spectroscopy \cite{foltynowicz2008noise}, in analysing atmosphere \cite{brown2003absorption} and chemical compositions \cite{nitkowski2008cavity,chen2016ultra,singh2019chemical}, or in medical diagnosis \cite{wojtas2015application}. Whilst conventional macroscopic Fabry-Perot cavities can be used to reach extreme sensitivity (analyte absorption coefficients of $<\SI{1e-9}{\per\centi\meter}$ \cite{berden2000cavity}), their susceptibility to ambient disturbance and their meter-scale size limits them to laboratory-based setups and requires large analyte volumes. This restricts their applicability under harsh external conditions or where a small-footprint device is required, like in aerospace technology or for highly integrated, low-maintenance industrial devices. Miniaturized resonator geometries on the other hand overcome these limitations, but usually face a trade-off between sensitivity and spectral coverage.

High finesse fiber Fabry-Perot cavities (FFPCs) \cite{hunger2010fiber,pfeifer2022achievements} have evolved as a miniaturized optical cavity platform within the last decade and have been used within numerous different experiments from interfacing single quantum emitters \cite{steiner2013single,gallego2018strong,albrecht2014narrow,toninelli2010scanning,besga2015polariton}, to mechanical resonators \cite{flowers2012fiber} or for various sensing tasks \cite{mader2015scanning,petrak2014purcell,wagner2018direct,mader2022quantitative}. They merge the advantages of Fabry-Perot cavities, an open mode volume and tunability of the cavity resonance, with the benefits of miniaturized resonators, a high level of integration, strong field-concentration and a small footprint. Although the smaller cavity length of FFPCs (mm-length record size \cite{ott2016millimeter}) impairs the reachable sensitivity in theses systems, the ability to produce high finesse resonators ($\mathcal{F} > \SI{100}{k}$, \cite{rochau2021dynamical}) partly compensates this drawback. This, together with their high level of miniaturization, the direct fiber integration and the large tunability, makes FFPCs superior to many competing platforms. 

In this article, we demonstrate gas sensing and spectroscopy of diatomic oxygen with a high finesse FFPC. We use the oxygen A-Band (about $\SI{755}{\nano\meter}$ to $\SI{770}{\nano\meter}$) to show the broadband performance of our devices that can resolve spectral fingerprints of molecules with a spectral resolution only limited by the employed laser source linewidth in a continuously sampled spectrum. Spectroscopy of oxygen absorption bands finds applications for example in the analysis of the structure and composition of earth's or even Mars' atmosphere \cite{mlynczak2001simultaneous,mlynczak2007sounding,guslyakova2014o2}. The operation range of our devices can be adjusted by using different cavity mirror coatings, for example into the near to mid infrared, to resolve the vibrational spectrum of atmospheric gases. We furthermore showcase a continuous, low-noise, high-speed scanning technique, which makes use of a modulation signal fed in a Pound-Drever-Hall (PDH) feedback loop. By using stable, tunable, high finesse FFPC, we demonstrate a miniaturized yet broadband and tunable spectroscopy device with high sensitivity and direct fiber coupling.

\section{Stable and tunable fiber cavity device}
\begin{figure*}
    \centering
    \includegraphics{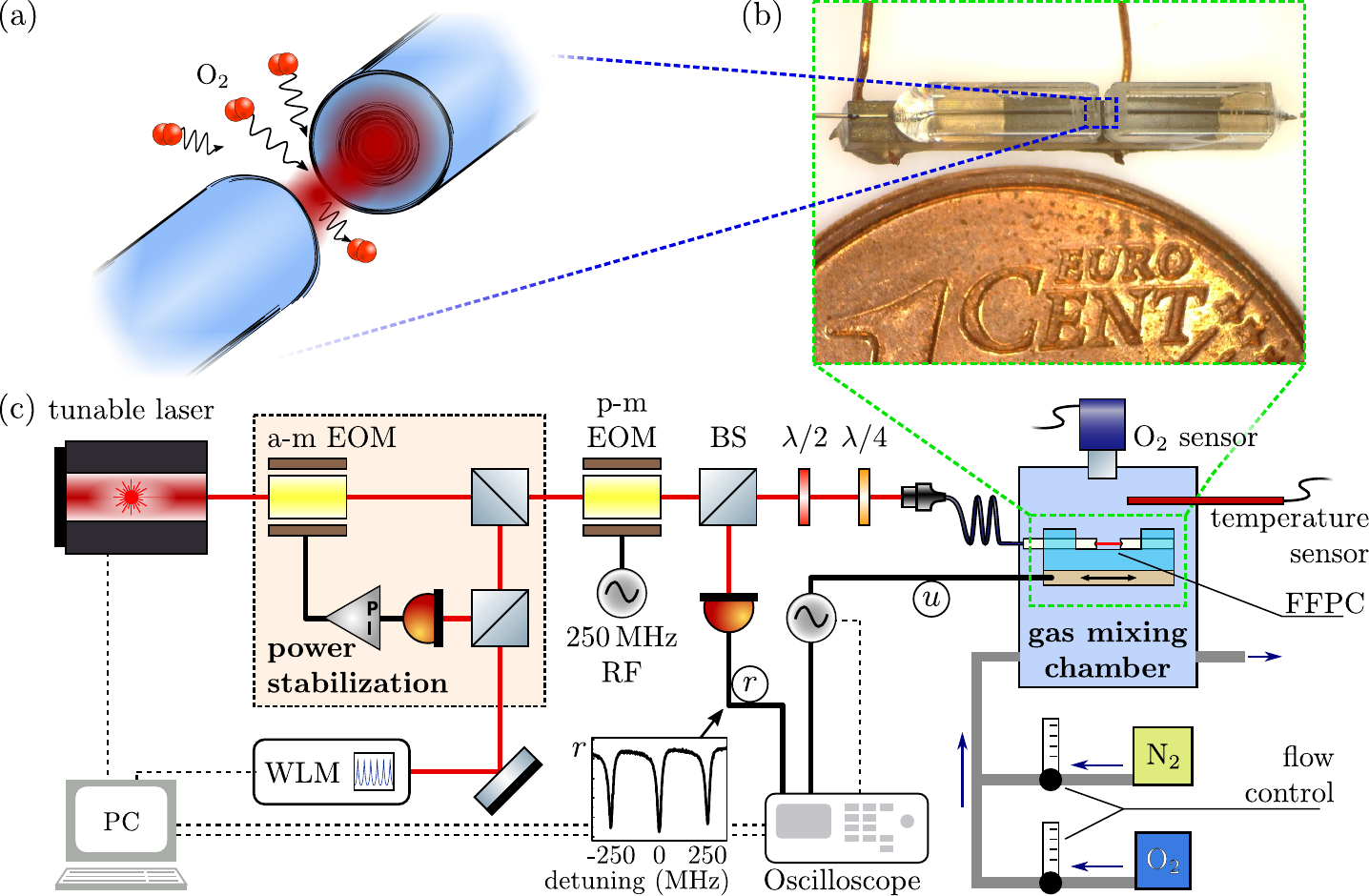}
    \caption{(a), schematic of \ce{O2} molecules entering into the mode volume of an FFPC. (b) picture of the FFPC device used in this article together with a \euro0.01 coin for size comparison. The measurement setup as used for the cavity mode width spectroscopy is depicted in (c). Here means: a-m/ p-m EOM -- amplitude/phase modulating electro-optic modulator, BS -- beam splitter, WLM -- wavelength meter, $\lambda/2$ and $\lambda/4$ -- half and quarter waveplate, PI is a servo loop, encircled sign shapes depict signal generators and half circles with black front depict photodiodes (sketch uses \cite{componenentLibraryInkscape}).}
    \label{fig:01-setupOverview}
\end{figure*}
The employed cavity design makes use of a recently introduced robust FFPC realization \cite{saavedra2021tunable}. It combines rigid, symmetric fiber Fabry-Perot cavities inside glass ferrules \cite{gallego2016high} with structural cuts and a piezo-electrical tuning element. Thereby, tuning over more than one free spectral range of the Fabry-Perot resonator is achieved. This FFPC design both features high tunability and passive frequency stability against ambient noise. The particular FFPC used in this article is shown in Fig.~\ref{fig:01-setupOverview}~(b). The two fiber mirrors (transmissivity: $ \mathcal{T} \approx \SI{20}{ppm}$, mirror losses $\mathcal{A} \approx \SI{26}{ppm}$, high reflectivity range: $\sim \SI{755}{\nano\meter} - \SI{815}{\nano\meter}$) are separated by about $\SI{93}{\micro\meter}$ constituting an FFPC with a base finesse $\mathcal{F}$ $\sim 67000$ at $\SI{762}{\nano\meter}$ (FWHM mode width: $\nu_\text{FWHM} \approx \SI{24}{\mega\hertz}$). Longer cavities, still with high finesse, are in principle possible for optimized mirror diameters $D$ and radii of curvature $ROC$ of the fiber mirrors, here: $(D,ROC) = (\SI{38}{\micro\meter}, \SI{170}{\micro\meter})$. 

The accessibility of the open mode volume of the FFPC allows to freely introduce different gas mixtures. In the presented experiment, we place the FFPC device inside a gas mixing chamber, where the concentration of oxygen at room temperature and ambient pressure is controlled by the flux rates of oxygen and nitrogen into the mixing chamber volume (see Fig.~\ref{fig:01-setupOverview}~(c)). For reference, the oxygen concentration in the chamber is monitored by an electrochemical oxygen sensor\footnote{Go Direct\textregistered O2 Gas (GDX-O2) from Vernier\texttrademark} and the inside temperature is measured to maintain reproducible experiment conditions. 

The FFPC system is used together with a widely tunable ($\sim \SI{745}{\nano\meter} - \SI{782}{\nano\meter}$) diode laser\footnote{Lion Series ECDL from Sacher Lasertechnik} (laser linewidth $<\SI{200}{\kilo\hertz}$ within $\SI{1}{\milli\second}$) with an additional power stabilization and wavelength readout\footnote{Wavelength Meter WS7-60 from HighFinesse} as shown in Fig.~\ref{fig:01-setupOverview}~(c). A phase-modulating electro-optical modulator (EOM) is incorporated after the laser power stabilization to add sidebands spaced by $\SI{250}{\mega\hertz}$. They are either used to increase the number of probed frequencies in a single cavity scan in the cavity mode width spectroscopy or for generating the error signal in the PDH-lock\cite{drever1983laser} of the PDH modulation spectroscopy. 

We monitor the reflection signal $r$ from the FFPC, while the cavity resonance is scanned over the laser using the attached piezo (drive signal $u$). The reflection signal is split from the incident laser beam using a non polarizing beam splitter (90/10). A combination of a half and quarter wave plate is used to couple light to one polarization mode of the cavity. The observed signal features three reflection dips at the EOM sideband spacing ($\SI{250}{\mega\hertz}$) from which the cavity mode width is extracted (FWHM empty cavity mode width $\nu_\text{FWHM}\approx \SI{24}{\mega\hertz}$).

\section{Oxygen absorption spectroscopy}
We test our spectroscopic sensor using the well-known oxygen A-band \cite{foltynowicz2008noise,greenblatt1990absorption,drouin2017multispectrum,van2004measurement,drouin2013high} that shows characteristic absorption features in the range from $\sim \SI{759}{\nano\meter} - \SI{770}{\nano\meter}$\footnote{All simulated spectrum data in this article is obtained using the web-based tool SpectraPlot: \href{https://www.spectraplot.com/}{https://www.spectraplot.com/} \cite{goldenstein2017spectraplot}, which makes use of NIST's HITRAN database\cite{rothman2013hitran2012}.}. The A-band manifold of vibration assisted transitions consists of absorption features with a collision- and Doppler-broadened FWHM linewidth of few GHz and maximum absorption coefficients of up to $\sim \SI{1.4e-3}{\per\centi\meter}$. Our results can however be transferred to other species and wavelength ranges for adjusted mirror coating parameters. To characterize the sensitivity of our device, we use the resolvable analyte absorption coefficient. In principle, the device can be used to investigate any other loss process caused by the analyte medium. 

\subsection{Cavity mode width spectroscopy}
\begin{figure}[ht!]
    \centering
    \includegraphics{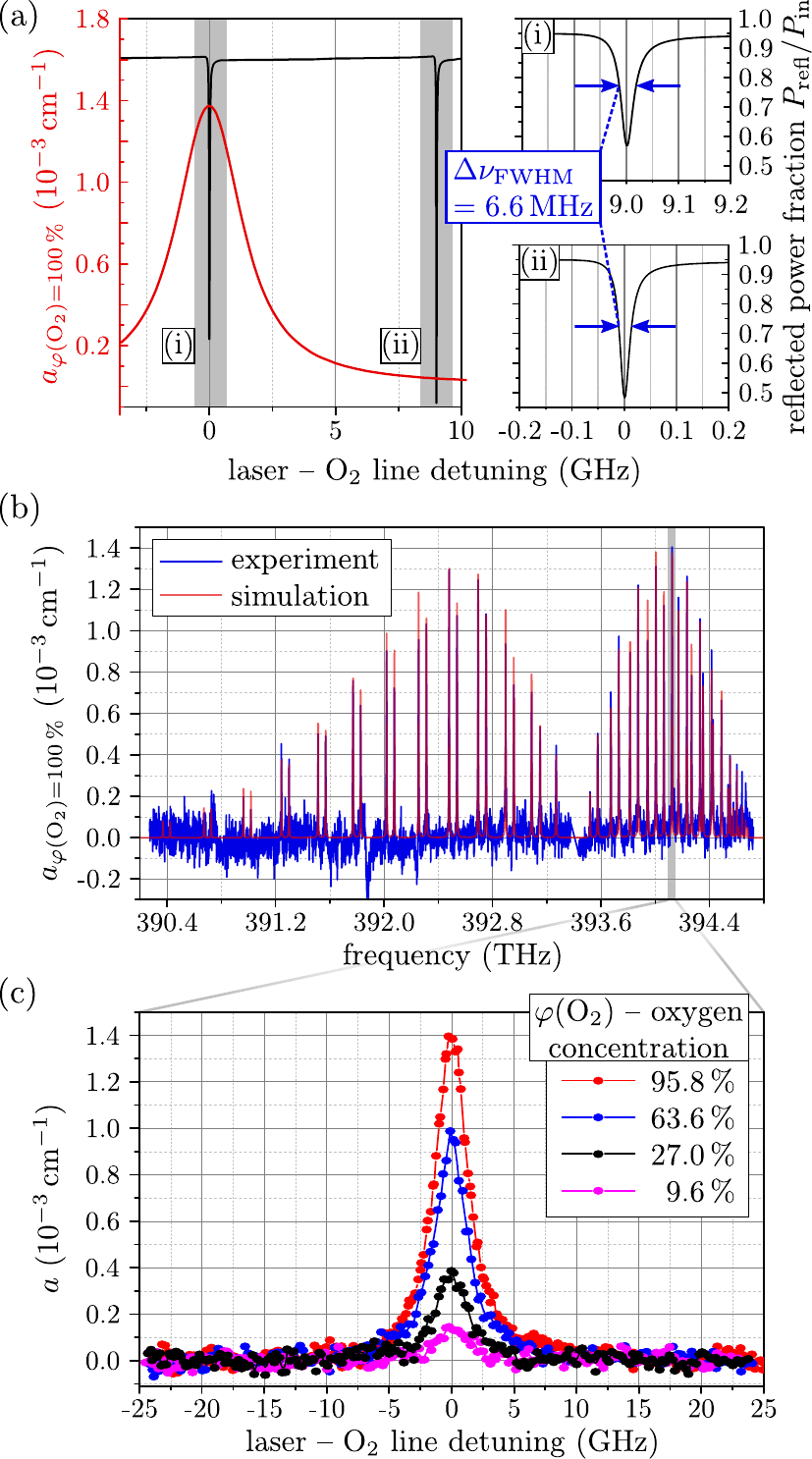}
    \caption{Schematic and measurements using cavity mode width spectroscopy (CMWS). The working principle of CMWS is illustrated in (a) for the typical parameters of our cavity. The additional absorption in the cavity by $\SI{100}{\percent}$ oxygen filling amounts to a change of $\Delta_\text{FWHM} = \SI{6.6}{\mega\hertz}$ of the cavity mode width compared to a spectrum point next to the absorption line. A full measurement of the $\SI{4.45}{\tera\hertz}$ broad spectrum of the oxygen A-band is shown in (b). The expected oxygen spectrum retrieved from SpectraPlot \cite{spectraplot, goldenstein2017spectraplot} is plotted for comparison. (c) shows measurements of a single absorption line at $\SI{394.1246}{\tera\hertz}$ for different volume concentrations of oxygen $\varphi(\ce{O2})$ (measured with a separate commercial \ce{O2} sensor) demonstrating the concentration sensing capability of the device.}
    \label{fig:02-CMWSdata}
\end{figure}

Cavity mode width spectroscopy\cite{cygan2013cavity} (CMWS) is a spectroscopy technique that uses the change of the linewidth of an optical resonator mode to determine the absorbance of a medium inside. For this, we express the finesse $\mathcal{F}$ of the Fabry-Perot resonator in terms of the resonator round-trip losses $\mathcal{F} = 2\pi/\left( 2\mathcal{M} + \sum_i \left( \mathcal{T}_i + \mathcal{A}_i  + \mathcal{C}_i\right) \right)$, where $\mathcal{T}_i$, $\mathcal{A}_i$, $\mathcal{C}_i$ are the transmission, absorption and clipping losses of the $i$-th mirror and $\mathcal{M}$ are the losses caused by the medium inside upon a single transmission. The cavity mode width directly relates to the finesse via the free spectral range $\nu_\text{FSR}$ of the cavity through $\mathcal{F}=\nu_\text{FSR}/\nu_\text{FWHM}$ with $\nu_\text{FSR} = c/2L_\text{cav}$, where $c$ is the speed of light and $L_\text{cav}$ is the cavity length. We obtain $\nu_\text{FSR}$ by measuring the cavity length and $\nu_\text{FWHM}$ is measured by scanning the cavity length such that successively the two sidebands and the main laser tone come into resonance with the cavity mode. The $\SI{250}{\mega\hertz}$ spacing of the laser tones is thereby used as a frequency meter for translating the measured resonance width into $\nu_\text{FWHM}$\cite{saavedra2021tunable}. $\sum_i \left( \mathcal{T}_i + \mathcal{A}_i  + \mathcal{C}_i\right)$ of the FFPC is known from measuring $\nu_\text{FWHM}$ with a cavity without absorbing medium, either by using an inert species inside or by measuring off-resonant to the absorption lines of oxygen. As the expected amount of absorption is on the order of parts per million, we linearly approximate Beer-Lambert's law such that $\mathcal{M} = a\cdot L_\text{cav}$, where $a$ is the absorption coefficient of the medium. Knowing $L_\text{cav}$ and the sum of mirror losses, $a$ can therefore be obtained directly by measuring $\nu_\text{FWHM}$. This basic CMWS scheme is also depicted in Fig.~\ref{fig:02-CMWSdata}~(a). Neglecting the effects of collision induced absorption (effects \ce{O2}-\ce{O2} versus \ce{O2}-\ce{N2} collisions \cite{greenblatt1990absorption,newnham1998visible}), the absorption coefficient is expected to be linearly dependent on the oxygen concentration at ambient conditions (room temperature and $\SI{1}{atm}$). The expected change of the optical mode-width $\Delta \nu_\text{FWHM}$ correspondingly reads 
\begin{equation*}
    \Delta \nu_\text{FWHM} = \frac{\nu_\text{FSR}}{2\pi} \cdot \mathcal{M} = \frac{ca}{4\pi}.
\end{equation*}

To determine the absorption coefficient at a given point in the spectrum, we set our laser to a specific investigation wavelength $\lambda_\text{inv} = c/\nu_\text{inv}$. We then tune the cavity resonance over the laser tone such that subsequently the first EOM-modulated sideband, the main laser tone, and the second sideband are getting into resonance. During the scan, the measured reflection signal $'r'$ features a train of three cavity dips (see inset of Fig.~\ref{fig:01-setupOverview}) with a Lorentzian plus dispersive shape\cite{gallego2016high,bick2016role}. The mode widths of the dips are extracted from fitting the measured lineshapes, yielding measurements at $\nu_\text{inv}$ and $\nu_\text{inv}\pm\SI{250}{\mega\hertz}$. The average mode width and standard deviation are obtained by tuning the cavity twenty times over the laser tones at each $\lambda_\text{inv}$. To measure the spectrum of an oxygen absorption line, $\lambda_\text{inv}$ is step-wise scanned over the feature with measurements between each step. Arbitrarily fine spaced $\lambda_\text{inv}$ with spectral information down to the line width of the laser are in principle possible at the cost of long measurement times. For wide spectral measurements like a molecular fingerprint a coarser spacing is used.

Using the described technique, we measured the full oxygen A-band between $\SI{759}{\nano\meter}$ to $\SI{769}{\nano\meter}$ as shown in Fig.~\ref{fig:02-CMWSdata}~(b). To optimize the resolution and required measurement time, the spacing of frequencies in the recorded spectrum between the center laser tones is swept. At the high frequency end of the $\SI{4.45}{\tera\hertz}$ broad spectrum the spacing is $\SI{500}{\mega\hertz}$ and at the low frequency end $\SI{1.3}{\giga\hertz}$. The acquisition of the full spectrum with twenty scans per investigated spectrum point was performed within approximately three hours. Faster acquisition of the spectrum either comes at the cost of less statistical information per spectrum point or less sampling density, as faster scanning of the cavity resonance has to be avoided to stay in the steady cavity field regime (for details see supplementary material part A). Since the optical properties of the coating slightly vary within the full spectrum range, the baseline mode width was measured and adjusted accordingly in the evaluation of the cavity spectra. The measured A-band lines are in good agreement with the expected spectrum within the expected absolute calibration error of the utilized wavelength meter ($\approx \SI{70}{\mega\hertz}$).

As the cavity resonance frequency is continuously tunable, the measurement resolution is only limited by the linewidth of the available laser source. This allows a high density sampling of individual absorption lines as shown in Fig.~\ref{fig:02-CMWSdata}~(c). At ambient conditions the absorption coefficient at the oxygen A-band lines is to very good approximation linear with the concentration of oxygen\cite{greenblatt1990absorption,newnham1998visible}. Therefore, the FFPC-device can also be used as a concentration sensor. The low concentration limit of this sensor is given by the accuracy of the mode width measurement. For CMWS, this accuracy is determined by the cavity frequency noise resulting from the integration of the noise spectrum over frequencies present in a cavity scan \cite{saavedra2021tunable} (see supplementary material part B). For the scan parameters used here, the inaccuracy amounts to $\Delta\nu_\text{FWHM} \approx \SI{167}{\kilo\hertz}$ corresponding to $\Delta a = (6.4\pm 0.2) \times 10^{-5}\SI{}{\per\centi\meter}$ or $\SI{4.6}{\percent}$ minimum oxygen concentration. This could potentially be reduced by improving the cavity frequency stability e.g. by a more stable FFPC design\cite{saavedra2021tunable}. Faster cavity scans that limit the influence of low frequency noise would not be suitable as they would not yield a steady state intra-cavity field (see supplementary material part A). Whilst CMWS in general is a straightforward measurement technique, it has a rather high sensitivity limit. It can be improved by using longer FFPCs, lower absorption mirror coatings or by changing to other operation modes like cavity ring-down spectroscopy \cite{berden2000cavity,zalicki1995cavity} (CRDS), phase-shift CRDS \cite{engeln1996phase,cheema2012simultaneous}, cavity-enhanced frequency-modulation spectroscopy \cite{ye1998ultrasensitive,ma1999ultrasensitive} or the closely related noise-immune cavity-enhanced optical heterodyne molecular spectroscopy (NICE-OHMS) \cite{foltynowicz2008noise,gianfrani1999cavity}. 

\subsection{PDH modulation spectroscopy}
\begin{figure*}
    \centering
    \includegraphics{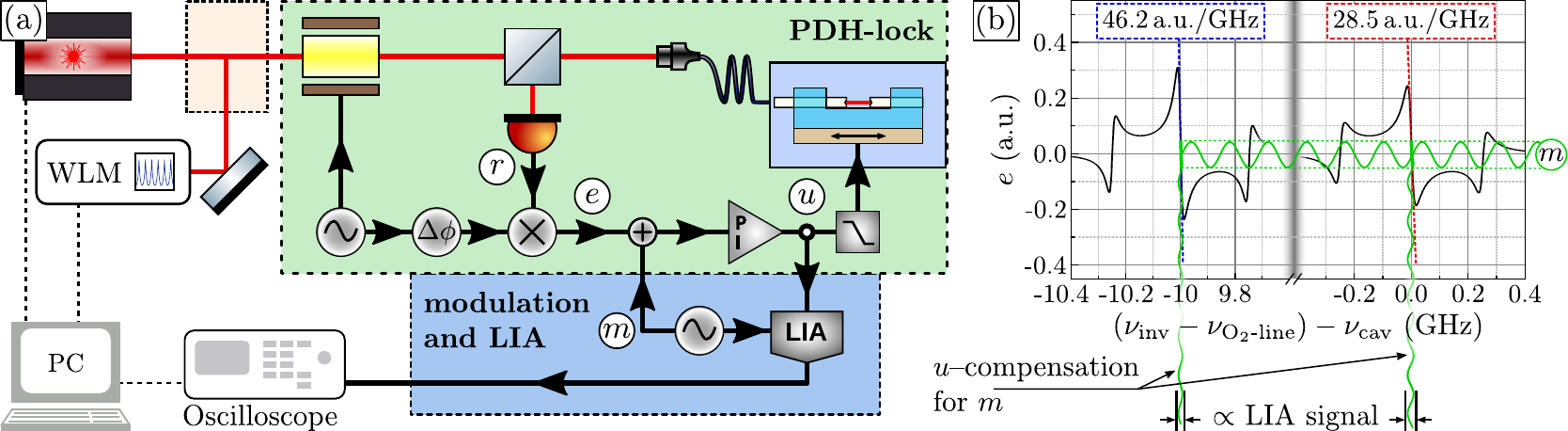}
    \caption{Schematic of the PDH modulation spectroscopy (PDH-MS). In (a) a simplified setup sketch for PDH-MS is shown. A modulation signal $m$ is added to the error signal $e$. This artificial perturbation is compensated by the feedback loop on $u$. This compensation is inversely proportional to the slope of $e$ at the lock point and is extracted using a lock-in amplifier (LIA). The exact relation between the slope and intra-cavity losses is detailed in supplementary material part C. (b) is a schematic showing the different signals employed. Other insertion and extraction points are possible depending on the frequency of $m$, the bandwidth of the cavity piezo, and the PI-controller.}
    \label{fig:03-PDHMSscheme}
\end{figure*}
Commonly used techniques like NICE-OHMS use a simultaneous excitation of cavity resonances by EOM-modulated sidebands of the laser. In the case of our $\SI{90}{\micro\meter}$ long cavity, this would require the generation of sidebands at several THz spacing. It is thus only applicable for longer FFPC realizations (e.g. with mm-long cavities \cite{ott2016millimeter} or hollow core fiber-based cavity geometries \cite{flannery2018fabry}). Nevertheless, one can use the interference signal of sidebands with less spacing to increase the sensitivity of the FFPC-enhanced spectroscopy. 
For this, we utilize the PDH error signal \cite{drever1983laser,black2001introduction} that is used to lock the cavity to the laser source. By inserting an additional modulation signal $m$ in the locked-cavity PDH-feedback loop, we measure the slope of the PDH error signal that is directly connected to the intra-cavity losses, thus coining this scheme PDH modulation spectroscopy (PDH-MS). The benefits of this method are a reduced noise background and a considerable increase of the measurement rate. A schematic of the measurement setup is shown in Fig.~\ref{fig:03-PDHMSscheme}. Here, we choose to add a modulation signal $m$ to the measured error signal before the PI-controller featuring a proportional (P) and integrator (I) control system part. The frequency of $m$ is below the feedback bandwidth of the system, thus the controller will act on the cavity system to actuate it such that the un-modulated error signal $e$ compensates for $m$ resulting in $m+e=0$. This requires a modulation of the driving signal $u$ of the cavity piezo at the modulation frequency with a magnitude depending on the slope of $e$. To extract this frequency component of $u$ and thereby $1/e'(\Delta\nu = 0) = 1/\left(\partial e /\partial \Delta\nu |_{\Delta\nu = 0}\right)$, with $\Delta\nu$ the drive detuning from the cavity resonance, we use a lock-in amplifier (LIA)\footnote{LIA-MV-200-H from FEMTO Messtechnik GmbH}. The LIA extracts and amplifies the component of $u$ at the modulation frequency. The cavity is thereby in a state locked to the laser and the output voltage of the LIA is proportional to $e'$. If the laser is scanned over an absorption line of oxygen, the cavity linewidth and mode matching change, resulting in a changed $e'$ that can directly be read off the signal from the LIA. For a full mathematical treatment of the dependence of the measured signal on the cavity mode width and the impedance matching of the cavity see supplementary material part C.

There are several possible options for setting up the modulation signal and readout within the feedback loop or by modulating the laser frequency. The particular choice mostly depends on the targeted modulation frequency. 
For example, for a modulation with a frequency larger than the feedback bandwidth of the PI-controller, but lower than the bandwidth of the piezo-element, $m$ can be added onto $u$ and then be read-off directly on $e$. The modulation signal should be chosen to fit in a low-noise area of the frequency-noise spectrum of the locked cavity \cite{saavedra2021tunable} thereby minimizing the measurement noise of the PDH-MS. In the presented scheme (see Fig.~\ref{fig:03-PDHMSscheme}), a modulation frequency of $\SI{2.2}{\kilo\hertz}$ below the PI-controller feedback bandwidth was used with an amplitude corresponding to $\Delta\nu < \SI{1}{\mega\hertz}$ well within the center linear slope part of the PDH error signal.

\begin{figure}[ht!]
    \centering
    \includegraphics{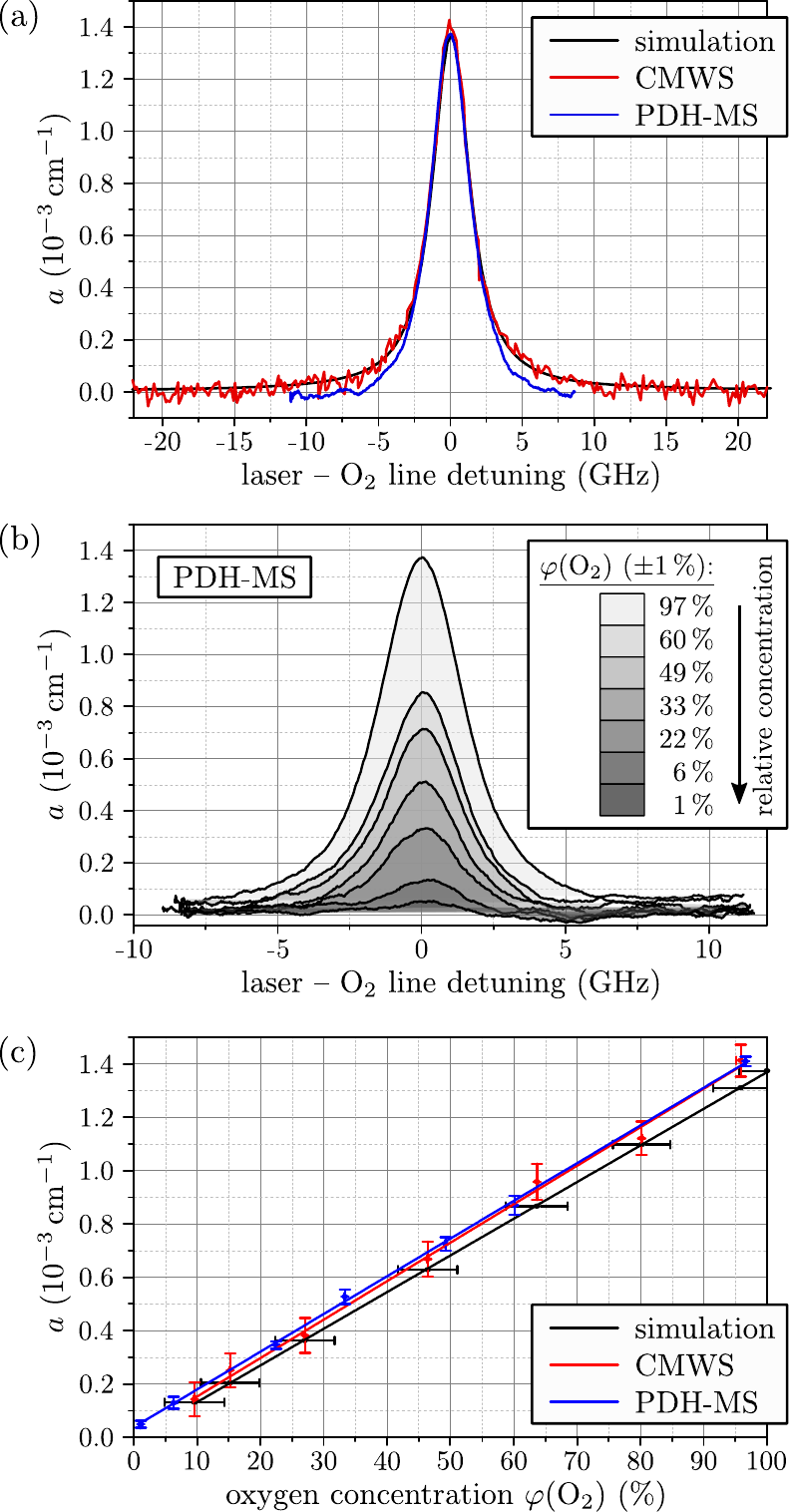}
    \caption{Performance of PDH modulations spectroscopy (PDH-MS) compared to cavity mode width spectroscopy (CMWS) (indicated simulation data obtained from SpectraPlot \cite{spectraplot, goldenstein2017spectraplot}). (a) shows a scan over a single absorption line at $\SI{394.1246}{\tera\hertz}$ using both PDH-MS and CMWS. For a discussion of the deviation of the PDH-MS signal at the far out slopes see supplementary material part C. In (b) PDH-MS scans for different oxygen concentrations (measured using the separate electro-chemical sensor) of the same absorption line are shown. In (c) the oxygen concentrations in different experiment runs are derived from both the absorption peak height measured with CMWS an PDH-MS and compared to the theoretical relation (simulation error bars indicate worst case systematic errors).}
    \label{fig:04-PDHMSdata}
\end{figure}

As the FFPC is tunable by more than one free spectral range, this measurement technique can be used for continuous, wide range frequency scans, provided the utilized laser source runs mode-hop free within the scan. Scanning a single absorption line of oxygen as shown in Fig.~\ref{fig:04-PDHMSdata}~(a) was performed within $\SI{10}{\second}$ compared to $\SI{20}{\minute}$ for the shown high-sampling rate line scans using CMWS, dramatically reducing the measurement time. The required scan time for a fixed frequency range using PDH-MS is determined by the laser scan speed, here: $\sim\SI{0.8}{\giga\hertz/\second}$, and the LIA integrator setting, here: $\SI{300}{\milli\second}$ integration time. The LIA integrator thereby sets the amount of averaging and has to be adjusted to be significantly below the time required for scanning the absorption feature and larger than several periods of the modulation signal frequency. Although it was not set up within the presented experiment, higher modulation frequencies can be beneficial in terms of the required scan time as well as for minimizing the measurement noise, if a low noise part of the frequency noise spectrum of the FFPC is picked. The advantage to use a sinusoidal signal for scanning the laser frequency is to avoid coupling higher frequency components of more complex signals to the mechanical modes of the resonator, which can result in locking instability (for example for triangular shapes).

Due to the single frequency readout of the LIA, the frequency noise of the cavity that otherwise affects the mode width measurement in the CMWS is strongly suppressed. The resulting resolvable absorption coefficient is reduced to $\Delta a = (2.2\pm 0.07) \times 10^{-5}\SI{}{\per\centi\meter}$ compared to $\Delta a = (6.4\pm0.2) \times 10^{-5}\SI{}{\per\centi\meter}$ for CMWS, also visible in the reduced noise of the PDH-MS signal in Fig.~\ref{fig:04-PDHMSdata}~(a). It can therefore be used to resolve lower concentrations down to $\SI{1}{\percent}$ as shown in Fig.~\ref{fig:04-PDHMSdata}~(b). More noise suppression for even higher sensitivity is possible for other insertion point schemes of the modulation signal and higher modulation frequencies as the cavity frequency noise in general decays for higher frequencies \cite{saavedra2021tunable}. 

\section{Conclusion}
In this article, we presented a fiber-based, miniaturized Fabry-Perot cavity that is used as a spectroscopy sensor for high resolution spectroscopy of the oxygen A-band as well as for measuring the concentration of oxygen down to one percent accuracy. The continuous, large tuning range of the cavity enables to record absorption spectra over tens of nanometers with a spectral resolution only limited by the linewidth of the employed light source. It can therefore serve to resolve spectral fingerprints of gases and at the same time be used to measure their concentration. Due to the miniaturized, light-weight and robust FFPC realization, it is suited for operation under demanding conditions like in aerospace or industry applications, where small analyte volumes have to be investigated. We furthermore demonstrated a low-noise technique for cavity enhanced absorption spectroscopy that features continuous fast scanning of the spectrum with improved sensitivity. It is especially suited for miniaturized cavities due to their large free spectral range that prohibits other high sensitivity techniques like NICE-OHMS \cite{gianfrani1999cavity}. 

There are numerous possible extensions and improvements to our proof-of-principle demonstration. FFPCs with adjusted coating parameters for longer wavelength operation can be directly used as miniaturized sensors for atmospheric trace gases. Higher sensitivity can be reached by longer FFPCs \cite{ott2016millimeter} or modified open cavity geometries with longer resonator length \cite{flannery2018fabry}. This will also enable the use of other established techniques like NICE-OHMS or different CRDS measurment schemes. Furthermore, alternatives like facilitating ring-downs introduced by a fast scanned cavity \cite{bresteau2017saturation} or thermal nonlinearities in the cavity may be employed, (see supplementary material parts A and D). These other techniques or modified versions of the demonstrated PDH modulation spectroscopy can further increase the sensitivity of the devices. As FFPCs easily operate in the fast cavity regime, they can also serve as a platform for photoacoustic spectroscopy \cite{west1983photoacoustic,wang2017fiber}. Finally, other areas of application are expected to open, if the FFPC are not used for investigating gaseous but liquid media. There are no limitations that exclude their combination with liquids, where they could serve as highly sensitive measurement tools for e.g. biological markers like dyes or as recently demonstrated for tracking nanoparticles \cite{kohler2021tracking}.

\begin{acknowledgments}
The authors thank Ryan M Briggs from Jet Propulsion Lab for the initial motivation of this project. The authors acknowledge funding by the Bundesministerium für Bildung und Forschung (BMBF) -- project: FaResQ, and funding by the Deutsche Forschungsgemeinschaft (DFG, German Research Foundation) under Germany's Excellence Strategy – Cluster of Excellence Matter and Light for Quantum Computing (ML4Q) EXC 2004/1 – 390534769. During the project course, C.S. was supported by a national scholarship from CONACYT, México. This work was supported by the Open Access Publication Fund of the University of Bonn.
\end{acknowledgments}

%%%%%%%%%%%%%%%%%%%%%%%%%%%%%%%%%%%%%%%%%
\clearpage
\appendix
\begin{widetext}
\begin{center}
    \large
    \textbf{Supplementary materials}
\end{center}
\end{widetext}
\renewcommand\thefigure{\thesection.\arabic{figure}} 
\renewcommand{\appendixname}{Supplementary material part}
\section{Swept cavity ring-down for fast scanned cavities}
\begin{figure}[htbp]
    \centering
    \includegraphics[scale=0.3]{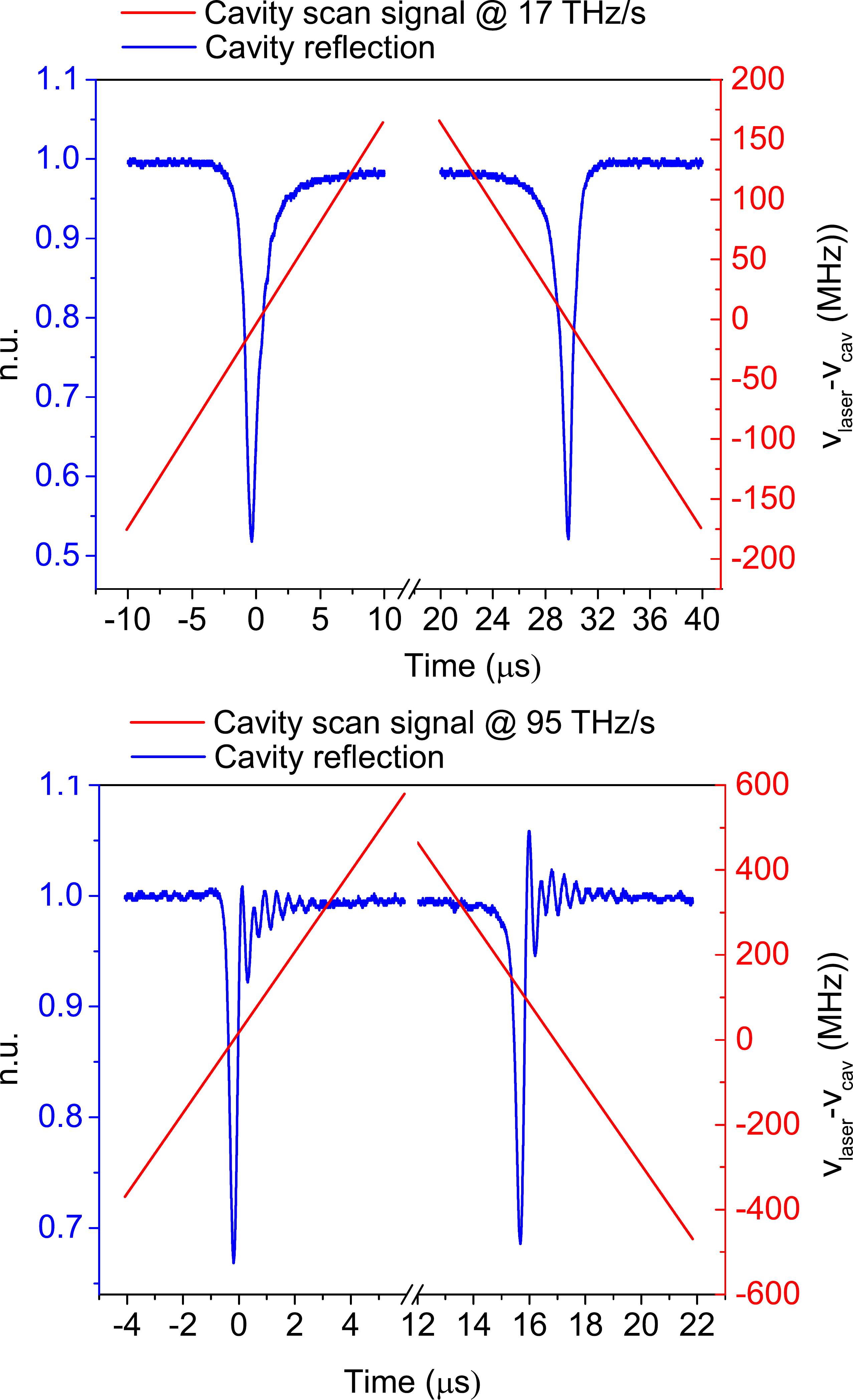}
    \caption{Comparison of quasistatic (top) and fast (bottom) cavity scans. The reflection in normalized units of an excitation laser with input power of \SI{25}{\micro\watt} to avoid nonlinear thermal effects is observed. The cavity mode width is approx $\SI{24.7}{\mega\hertz}$ and the two scan speeds $\SI{17}{\tera\hertz/\second}$ and $\SI{95}{\tera\hertz/\second}$. Fast acquisition of data using CMWS is limited by the ringing effect at the latter high speed scan.}
    \label{chirp}
\end{figure}

To measure the cavity mode width, we scan the cavity length at a fixed laser wavelength. Sidebands at $\SI{250}{\mega\hertz}$ created by an electrooptic modulator are used as frequency markers to calibrate the scan. For consistent measurements of the cavity mode width, the intracavity field has to be in a steady state with the excitation laser. This is achieved, if the cavity is scanned in a quasistatic way, meaning that the time $T$ required for scanning over the resonance is much bigger than the rate at which the cavity is coupled to the environment: $T\gg1/\Delta \nu_\text{FWHM}$. Here, we reach this regime, when the scan speed is maintained at a few THz/s. Figure~\ref{chirp} shows the influence of the scan speed on the measured trace. If faster scans are used, the reflected light and the cavity field interfere and a ringing is observed as light is still leaking out of the cavity, while the laser is already off-resonant. The ringing has an exponential envelope connected to the cavity decay and a slight chirp as the cavity field is frequency shifted by the Doppler effect from the moving mirrors. The raised/ lowered energy of the photons thereby corresponds to the work done/ extracted by the moving mirrors being displaced against the light field's radiation pressure. 

Whilst the ringing effect limits the maximum scan speed for the data acquisition in the cavity modewidth spectroscopy (CMWS), it can itself be a valuable resource as an own characterization method for the cavity mode width. For this, one can either use a model as described in \cite{bresteau2017saturation} or simulate the cavity reflection for a swept pump. The cavity field $a(t)$ can for example be numerically computed using its equation of motion. In a frame rotating at the excitation laser frequency, this can be written as
\begin{align*}
\dot{a}(t) = - \left( iv_\text{scan}\cdot t +\frac{\kappa}{2} \right) a(t) -\sqrt{\kappa_e}\,a_\text{input}\;,
\end{align*}
where $v_\text{scan}$ is the scan speed of the cavity (in ($2\pi \cdot $ THz)/s), $\kappa$ is the full cavity energy decay rate and $\kappa_e$ the coupling rate to the excitation field $a_\text{input}$. The time steps for a numerical evaluation have to be chosen to lead to a convergent behavior, still reflecting the fast scan process compared to the the energy decay of the cavity. Exact fitting to this model is however complicated by the detuning dependent mode matching coupling to fiber-cladding modes in case of fiber Fabry-Perot cavities \cite{gallego2016high,bick2016role}. 

\section{Averaging and error for CMWS measurement}
\begin{figure*}[htbp]
    \centering
    \includegraphics[scale=0.9]{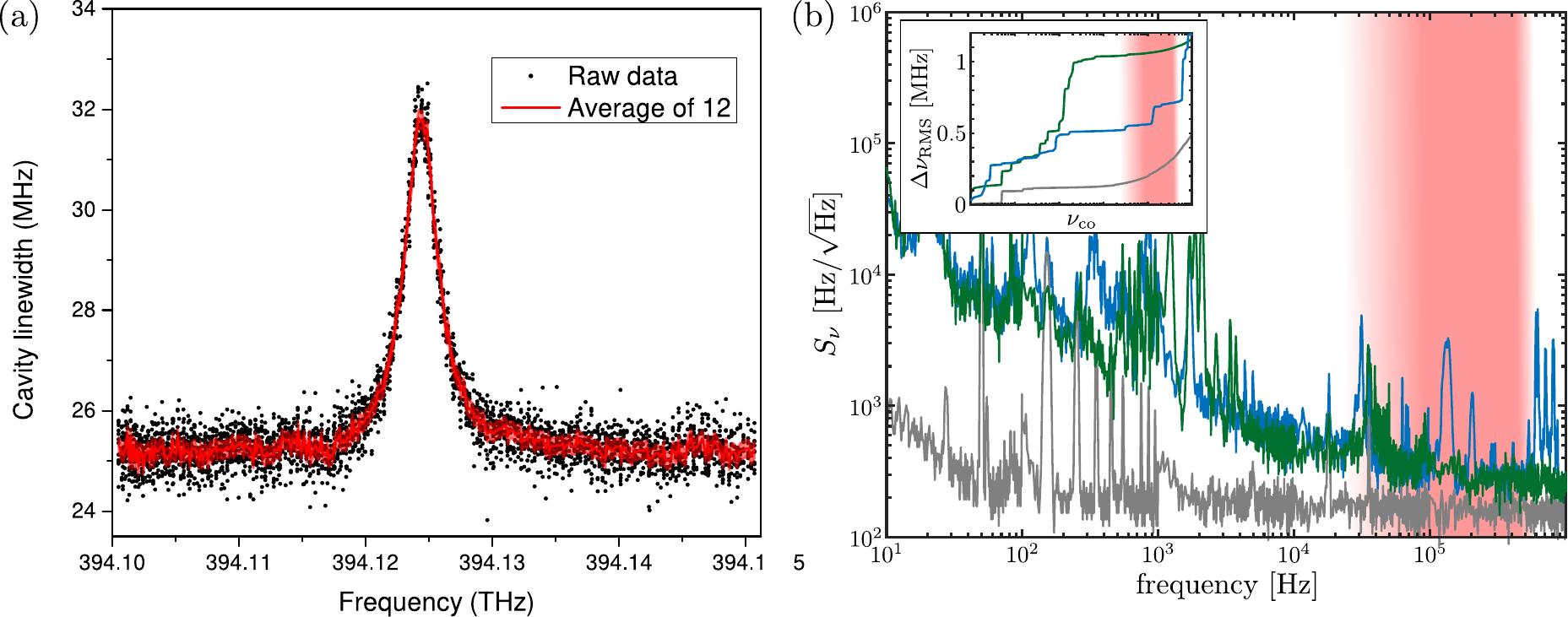}
    \caption{(a) Cavity mode width measurement across oxygen absorption line at 95\% concentration. The black dots each indicate a mode width measurement, here with 12 independent measurements per frequency. The red line represents the corresponding average and the shading the final statistic uncertainty. In (b) the cavity frequency noise spectrum of two different cavity designs (widely tunable \textit{single and triple slot cavities} as used in this article -- see supplementary of \cite{saavedra2021tunable}) is shown in green and blue (electrical background noise in gray). The fluctuation of the measured mode width is predominantly caused by noise in the red shaded area (some tens of $ \SI{}{\kilo\hertz} \rightarrow \SI{500}{\kilo\hertz}$). Slower noise does not play a role as the cavity scan is too fast for these noise processes to contribute. Higher frequency noise is avoided by low-pass filtering the oscilloscope input (low pass $\SI{3}{\dB}$ at $\SI{500}{\kilo\hertz}$). The inset shows the fully integrated noise (from 0 to $\nu_\text{co}$). For estimating the mode width noise only the steps of the integrated rms-noise within the red shaded area contribute.}
    \label{E}
\end{figure*}

The error and low concentration measurement limit of our CMWS measurements is caused by fluctuations in the cavity mode width measurement. To yield the data of the main manuscript 20 mode width measurements per frequency point were taken. The reported absorption coefficient for each frequency point corresponds to the statistic average of these measurements. The fluctuation of the measured mode width causes the uncertainty and lowest detectable amount of absorption. An exemplary set of data (here with only 12 measurements per point) is shown in Fig.~\ref{E}~(a). 

The fluctuations that limit the measurement accuracy are mostly caused by frequency noise of the cavity resonance from thermally excited high frequency mechanical resonances of the system within the red shaded area in Fig.~\ref{E}~(b)\cite{saavedra2021tunable} -- the exact contributions of each frequency can be determined by a Fourier transform of the measured signal. Lower frequency noise does not contribute within the scan time and noise at higher frequency is filtered by a low pass filter before the oscilloscope. The performance of FFPCs for CMWS measurements can potentially be strongly improved by FFPC designs that avoid mechanical resonances in the relevant frequency range.

\section{The influence of mode width and impedance matching in PDH modulation spectroscopy}
\begin{figure}[htbp]
    \centering
    \includegraphics[width=\columnwidth]{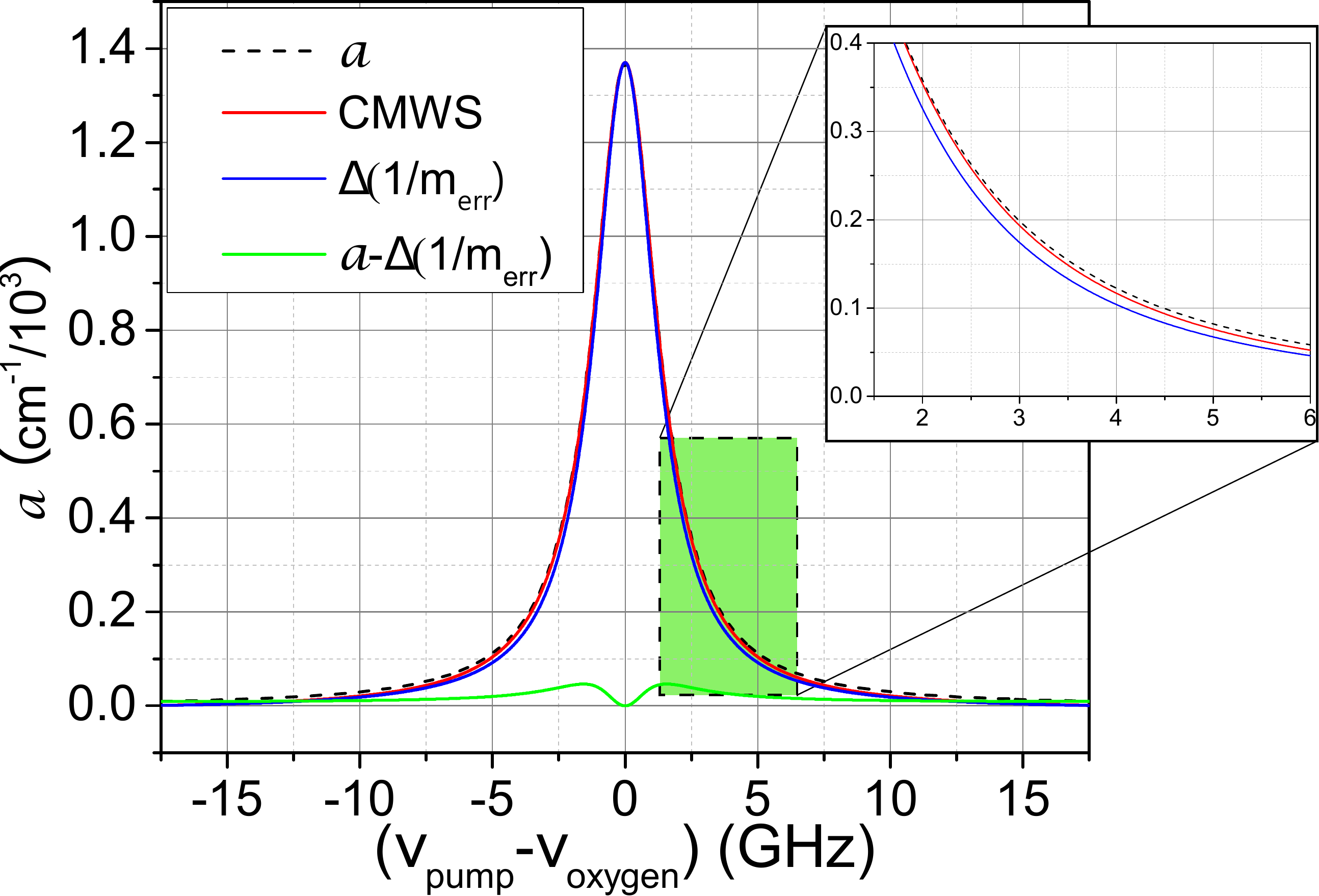}
    \caption{Comparison between PDH-MS and CMWS in the measured absorption coefficient on an exemplary oxygen line. A slight deviation of PDH-MS and CMWS is expected on the tails of the absorption line due to the modified impedance matching caused by the absorption.}
    \label{MS}
\end{figure}

Due to the absorption of light within the cavity, both the cavity mode width and impedance matching are changed affecting both Lorentzian width and depth of the reflection signal $\mathcal{R}$. This changes the Pound-Drever-Hall (PDH) error signal $e$ that depends on the cavity reflection signal via \cite{black2001introduction}
\begin{equation}
    e(\Delta\nu)=A\cdot \Im[\mathcal{R}(\Delta\nu)\mathcal{R}^*(\Delta\nu+\Omega)-\mathcal{R}^*(\Delta\nu)\mathcal{R}(\Delta\nu-\Omega)] \; ,
    \label{err}
\end{equation}
where the detuning $\Delta\nu$ is given by $\Delta\nu= (\nu_\text{pump}- \nu_\text{cav}) /\nu_\text{FSR}$ ($\nu_\text{pump}$ being the laser frequency, $\nu_\text{cav}$ the cavity resonance frequency and $\nu_\text{FSR}$ the cavity's free spectral range). $\Omega\gg \nu_\text{FSR}/\mathcal{F}$ is the modulation frequency of the employed electrooptic modulator (EOM) -- $\mathcal{F}$ being the cavity finesse. The proportionality constant $A$ is given by the product of the gain of the photodetector $G$ (V/W) times the amplitude of the modulated sidebands and the cavity pump power $A=2GJ_0(\beta_m)J_1(\beta_m)P_\text{pump}$ \cite{black2001introduction}, where $J$ denotes the Bessel function and $\beta_m$ the modulation constant for the EOM. We find that while locked at the cavity resonance, the inverse of the slope of $e$ at $\Delta\nu=0$ is proportional to the absorption coefficient of the absorption species within the cavity. As in \cite{gallegoThesis,gallego2016high}, we write the complex reflectivity as
\begin{equation*}
    \mathcal{R}(\Delta\nu)=\beta-\xi(\Delta\nu)\alpha^2(\Delta\nu) \; .
    \label{Re}
\end{equation*}
Here, $\alpha$, and $\beta$ denote the mode overlap of fiber and cavity, and fiber and reflected mode respectively. $\xi$ is the complex leakage ratio function of the FFPC. For sufficiently small cavity detuning ($\Delta\nu\approx 0$), the reflectivity of the modulated sidebands can be approximated as $R(\omega\pm\Omega)\approx\beta$. Using this, the measured error signal near resonance is
\begin{equation}
   e(\Delta\nu)\approx2A\cdot \Im\left[\mathcal{R}(\Delta\nu)\cdot\beta\right] \; ,
   \label{elin}
\end{equation}
and the FFPC complex leakage ratio function can be expressed as
\begin{equation*}
    \xi(\Delta\nu)\approx\frac{t^2re^{-aL}(1-r^2e^{-aL})+it^2re^{-aL}\pi\Delta\nu/FSR}{(1-r^2e^{-aL})^2} \; .
    \label{R}
\end{equation*}
$L$ thereby denotes the cavity length, $a$ the absorption coefficient of the intracavity medium, $t$ and $r$ the mirrors' dielectric transmissivity and reflectivity for the electric field. The overlap integrals $\alpha$ and $\beta$ expressed for small cavity detuning are \cite{gallego2016high} 
\begin{align*}
    \alpha&=\frac{4RW_{f}W_{m}[2R(W_{m}^2+W_{f}^2)+iW_{f}^2 W_{m}^2 k_f]}{4R^2(W_{m}^2+ W_{f}^2) +W_f^4W_m^4k_f^2} \\
\text{and } \qquad    \beta&=R\frac{4R-2iW_{f}^2k_{f}}{4R^2+W_{f}^4k_{f}^2} \; .
\label{overlap}
\end{align*}
Here, $W_{f}$ is the fiber mode field radius ($\sim\SI{2.5}{\micro\meter}$ for SM fiber), $W_m$  is the cavity mode waist at mirror interface with radius of curvature $R$ ($W_m=cR\sqrt{L}/(\pi\nu_\text{pump}\sqrt{2L-R})$, and $c$ the speed of light in vacuum. Both mirrors are assumed identical here. Their reflection coefficient (usually expressed in ppm) is given by the modulus square of the dielectric reflectivity $|r|^2=1-\mathcal{T}-\mathcal{L}$, $\mathcal{T}$ is the corresponding power transmissivity and $\mathcal{L}$ the mirror losses. Equation~\ref{elin} is plotted in the main text in Fig.~3~(b). Considering that the absorption losses due to the oxygen presence are much smaller than the mirror losses per round trip, we can extract the slope of the error function $m_e$, which is proportional to the absorption coefficient $a$
\begin{equation}
    1/m_e=\frac{\nu_\text{FSR}}{A}\frac{((1-r^4)aL+(1-r^2)^2)(4g^2+W_f^4W_m^4k_f^2)^2}{4\pi t^2rb(4g(\beta_iW_m^2W_f^2k_f-\beta_rg)+\beta_rW_m^4W_f^4k_f^2)}
    \label{slope2}
\end{equation}
Here, $\beta_{r,i}$ defines the real and imaginary parts respectively and  $b=(4RW_mW_f)^2$ and $g=R(W_m^2+W_f^2)$.  Finally, we obtain for the relative difference with respect to the absence of an absorbing gas species (here $m_{\text{N}_2}$)
\begin{equation}
    \Delta(1/m)=1/m_{\text{O}_2}-1/m_{\text{N}_2}=C\cdot a \; ,
    \label{slope3}
\end{equation}
where $C$ is a proportionality constant extracted from Eq.~\ref{slope2}. Equation~\ref{slope3} is plotted using the measured cavity geometry and optical parameters, as well as the absorption coefficient from SpectraPlot\cite{goldenstein2017spectraplot,spectraplot} in Fig.~\ref{MS} to show the equivalence of this technique and mode with spectroscopy. However, due to the effect of the changing impedance matching of the FFPC a slight deviation from the CMWS technique is expected on the wings of the absorption lines.

\section{Thermal nonlinearities in high power cavity scans}
\begin{figure}[htbp]
    \centering
    \includegraphics[scale=0.35]{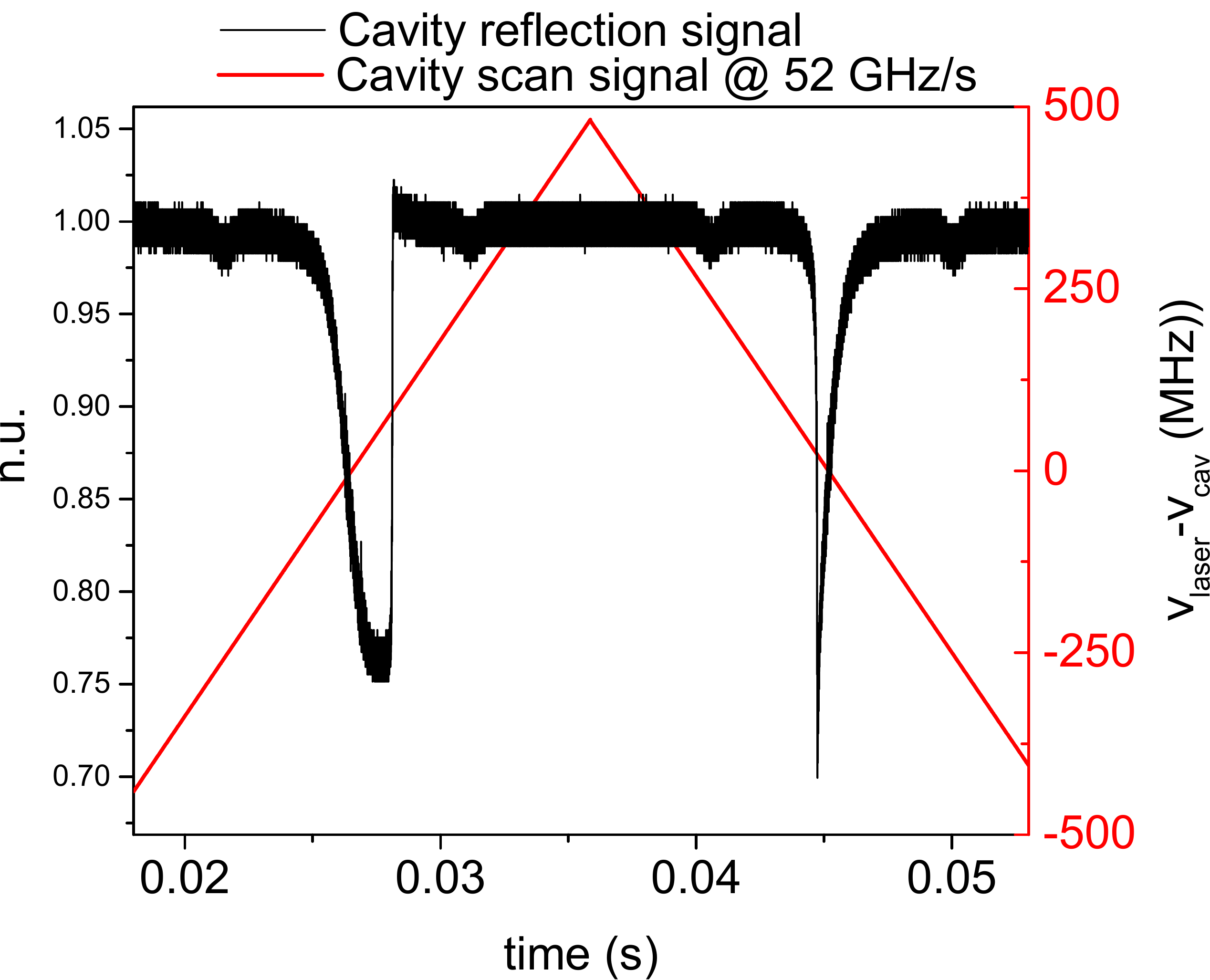}
    \caption{Cavity reflection signal with sidebands at 250 MHz measured using slow scan speeds and a pump power of \SI{310}{\micro\watt}. Depending on the scan direction, the cavity reflection signal becomes broader or narrower as the resonance is thermally shifted towards or away from the approaching laser tone. In both cases the shape is no longer a Lorentzian profile.}
    \label{T}
\end{figure}

 Nonlinear thermal effects \cite{brachmann2016photothermal} occur when the cavity resonance is scanned using a higher power laser input (e.g. some hundred $\SI{}{\micro\watt}$ compared to few tens of $\SI{}{\micro\watt}$ under normal operation) and slow -- in the time scale of thermal effects -- scan speeds. They are caused by an expansion of the fiber mirror material as a consequence of absorbed intracavity photons and manifest in a broadening and narrowing (see Fig.~\ref{T}) of the scanned line together with a deformation of the Lorentzian lineshape depending on the scan direction. For very high powers, a bistable regime of the cavity is reached \cite{carmon2004dynamical}. The deformed lineshape prohibits a correct cavity linewidth analysis and therefore can not be used for CMWS. It can however be avoided by using not too slow scan speeds below the swept cavity rings discussed in part A. 
 
%%%%%%%%%%%%%%%%%%%%%%%%%%%%%%%%%%%%%%%%%
\clearpage
\bibliographystyle{IEEEtran}
\bibliography{references}

\end{document}